\documentclass[twocolumn,prl,superscriptaddress,showkeys,showpacs]{revtex4}
\usepackage{amsmath,graphicx,amsfonts,bm,amssymb}
\usepackage{times}
\begin{document}
\title{Detecting non-Abelian geometric phase in circuit QED}

\author{Man-Lv Peng}
\affiliation{Laboratory of Quantum Information Technology, and
School of Physics and Telecommunication Engineering, South China
Normal University, Guangzhou 510006,  China}

\author{Jian Zhou}

\affiliation{Anhui Xinhua University, Hefei, 230088, China}

\affiliation{Laboratory of Quantum Information Technology, and
School of Physics and Telecommunication Engineering, South China
Normal University, Guangzhou 510006,  China}

\author{Zheng-Yuan Xue}

\affiliation{Laboratory of Quantum Information Technology, and
School of Physics and Telecommunication Engineering, South China
Normal University, Guangzhou 510006,  China}

\date{\today}

\begin{abstract}
We propose a scheme for detecting noncommutative feature of the non-Abelian geometric phase in circuit QED,
which involves three transmon qubits capacitively coupled to an one-dimensional transmission line resonator.
By controlling the external magnetic flux of the transmon qubits, we can obtain an effective tripod interaction of our
circuit QED setup. The noncommutative feature  of the non-Abelian geometric phase is manifested that for an initial
state undergo two specific loops in different order will result in different final states.
Our numerical calculations show  that this difference can be  unambiguously detected in the proposed system.

\end{abstract}



\maketitle


Phase factor has played a profound role in quantum physics.
Apart from the familiar dynamical phase, geometric phase (GP), discovered
 in 1984 by Berry \cite{Berry}, has deep
physical meanings. Berry pointed out that GP occurs when a system
is subjected to a cyclic adiabatic evolution,  which results from the geometrical
properties of the parameter space of the Hamiltonian.
Especially,  Wilczek and Zee found non-Abelian gauge phase
results from Berry's formula 
in 1984 \cite{FZ}. GP depends only on the solid angle
enclosed by the parameter path, and thus is robust against local
noises   \cite{AIM,GD,AIMV,PPN,IFE,SJY,zhuerror}. Therefore, it has been proposed
to implement fault-tolerant quantum logical gates for universal quantum computation
\cite{Zanardi, Pachos, Jones, Duan,zhuunconventional,xue}.

Non-Abelian GP  differs from Abelian geometric phase according to the commutation
features of their gauge potential, i.e., an initial state undergoes two specific cyclical evolution  in different order
will result in different final states. Up to now, Abelian GP has
been experimentally detected  in various systems
\cite{Tycko,e,f}, while non-Abelian GP has not been verified yet.
Usually, dark states are used in detecting GP so that dynamical phase will not appear.
Conventionally, it is the tripod Hamiltonian that has been proposed to detect the
non-Abelian GP \cite{RBK,Zhang,JGP}.
However, it is difficult to find tripod
configuration atomic energy levels, which
impedes experimental detection of the non-Abelian GP.
Recently, it is also proposed that non-Abelian GP can be detected by two laser beams
interacting with a three-level $\Lambda$ atom in cold atomic
system \cite{Du,eric}. Since two of the eigenstates in this scheme are only
near degenerate, dynamical phases will also occur during the process,
and thus one needs additional effort to conceal it \cite{Du}.
Meanwhile,  the non-Abelian GP is also proposed to be detected in a new designed multi-level
superconducting circuit \cite{feng}. However, multi-level scenario of this
superconducting nanocircuit is very sensitive to its background charge noise.
Therefore, to certify the fundamental non-Abelian nature of the non-Abelian GP,
it is of great importance to find an experimentally accessible system that can host the exotic non-Abelian structure.

Superconducting system is regarded as one of the most
promising candidates for physical implementation of qubits
which can support scalable quantum information processing \cite{you,nori}.
Furthermore, by placing superconducting qubits in a cavity, i.e,
circuit QED  setup \cite{cqed,ARA}, the system will have several practical
advantages including strong coupling strength, immunity to
noises, and suppression of spontaneous emission.
Here, we propose to detect the noncommutative feature of non-Abelian GP with effective tripod Hamiltonian in circuit QED.
The setup we consider  consists three transmon qubits  that are capacitively coupled to an
one-dimensional (1D) high-Q transmission line resonator (cavity), which has recently been realized
experimentally \cite{exp}. With proper chosen parameters, such setup can be effectively
described by the  tripod Hamiltonian  \cite{tripod}, and thus can
be used to  detect the noncommutative feature of non-Abelian GP. Furthermore, the transmon qubit possesses remarkable superiority \cite{JTJ},
e.g., it achieves exponential insensitivity to charge
noise without increasing the sensitivity to either
flux or critical-current noise.
Note that when adding a shunt capacitor to a flux qubit will also lead to low-decoherence qubit \cite{t2}.

\begin{figure}[tbp]
\includegraphics[width=8cm]{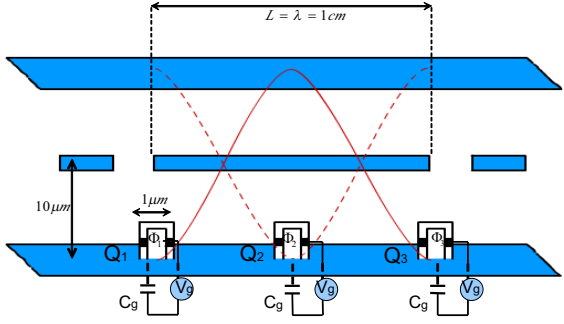}
\caption{(Color online) Schematic of the circuit QED
architecture. The 1D superconducting line resonator consists of
a full-wave section ( $L=\lambda=1$cm ) of superconducting coplanar
waveguide. Qubits placed between superconducting line resonators
consist of two small  Josephson junctions, which permit tuning of the
effective Josephson energy by an external flux $\Phi$ in a 1 ${\mu}$m
loop. There is approximate 10 ${\mu}$m gap between superconducting
line resonators. The input and output signals can be coupled to the
resonator via the capacitive gaps in the center line.}
\label{xitong}
\end{figure}

The considered  circuit QED architecture is shown in Fig. \ref{xitong} with three identical transmon qubits
that are  capacitively coupled to the cavity.
The transmon qubit has  effective Josephson energy
$E_{J}=E_{J, max}|{\cos(\pi\Phi/\Phi_{0})}|$  with $E_{J, max}$,  $\Phi$ and  $\Phi_{0}$ being the Josephson energy  of the Josephson junctions,
the external magnetic flux and the flux quantum,  respectively.
This type of qubit has good coherence performance.  The charging energy of the transmon
is much small compared with the Josephson energy ($E_{C}{\ll}E_{J}$).
With  $E_{J}/E_{C}=50$ ($E_{C}=0.3 $ GHz, $E_{J}=15$ GHz), the energy difference of the two lowest levels
(defined as first excited state $|e\rangle$ and ground state $|g\rangle$) is approximately $\sqrt{8E_{J}E_{C}}$,
and the relaxation time for $|e\rangle$  is on the order of 0.06 s \cite{JTJ}. For an 1D cavity with length $L=\lambda=1$ cm,
we can get  rms voltage $V^{0}_{rms}=\sqrt{\hbar\omega/lc}$ of an antinode between two  superconducting lines,
where $l$ and $c$ are the inductance and capacitance per unit length, respectively. As a result,
qubits are coupled to the superconducting line by means of the voltage $\hat{V}=V^{0}_{rms}(\hat{a}+\hat{a}^{\dag})$.
Remarkably, for  coplanar waveguide cavity, cavity quality factor $Q \sim 10^6$ has already been demonstrated \cite{PHB},
which means that the internal losses can be very low.
With three qubits fabricated at the antinodes of the cavity voltage, the strength of the coupling to
the resonator is maximized for all three transmom qubits. Then, the system can be described by the Tavis-Cummings Hamiltonian
 \begin{equation}
\label{Hamiltonian} \hat{H}_{TC}=\hbar\omega\hat{a}^{\dag}\hat{a}+\sum_{i=1}^{3}{\left[{1\over2}\hbar\varepsilon_{i}\hat{\sigma}_{z}^{i}+{\hbar}g_{i}
(\hat{\sigma}_{+}^{i}\hat{a}+\hat{\sigma}_{-}^{i}\hat{a}^{\dag})\right]},
\end{equation}
where $\omega=1/\sqrt{lc}$ is the resonator frequency, $\hat{a}$ ($\hat{a}^{\dag}$) is annihilation
(creation) operator of the 1D cavity mode,  $\varepsilon_{i}=\sqrt{8E_{c}^{i}E_{J}^{i}}/\hbar$  is the energy splitting of the $i$th qubit,
$\hat{\sigma}^{i}$ are Pauli operator for the $i$th transmon,  and $g_{i}=2eC_{g}^{i}V^{0}_{rms}(E_{J}^{i}/8E_{_C}^{i})^{1/4}/
{\sqrt{2}C_{_{\Sigma}}^{i}}$ is the strength of coupling between the $i$th qubit and the superconducting line.
For $C_g/C_{_{\Sigma}}=0.1$ with $C_{_{\sum}}=C_{g}+2C_{J}$, the coupling strength controlled by external
magnetic flux $\Phi_i$ will be on the order of 100 MHz \cite{ARA}. Describe $\varepsilon_{i}$ and $g_{i}$
of Hamiltonian in Eq. (\ref{Hamiltonian}) in external magnetic flux $\Phi_{i}$, they read
\begin{subequations}
\begin{equation}
\label{varepsilon}
\hbar\varepsilon_{i}(\Phi_{i})=\sqrt{ 8E^{i}_{C}E^{i}_{J,max}|\cos(\frac{\pi\Phi_{i}}{\Phi_0})|},
\end{equation}
\begin{equation}
\label{g}
g_i(\Phi_i)=\sqrt{2e^2}\frac{C_{g}^{i}}{C_{\Sigma}^{i}}V^{0}_{rms}(\frac{E_{J,max}^{i}}{8E_{C}^{i}})^{1/4}|(\cos\frac{\pi\Phi_{i}}{\Phi_0})|^{1/4}.
\end{equation}
\end{subequations}
Explicitly, the coupling strength $g_i$  and the qubit frequency  $\varepsilon_i$  are endowed with a relation $g_i(\Phi_i)\varpropto\sqrt{\varepsilon_i(\Phi_{i})}$.

Restrict the system in the only one-excitation subspace of the dynamics states {$\{|1ggg\rangle,|0egg\rangle,|0geg\rangle,|0gge\rangle\}$},
where $|0\rangle$ and $|1\rangle$  denote the cavity mode has 0 and 1 microwave photon. Then,  in this subspace,
the  interaction Hamiltonian (\ref{Hamiltonian})  can be written as
\begin{equation}
\label{interaction}
\hat{H}_{sub}=\begin{bmatrix}
0&g_1&g_2&g_3\\g_1&\Delta_1&0&0\\g_2&0&\Delta_2&0\\g_3&0&0&\Delta_3
\end{bmatrix},
\end{equation}
where $\Delta_i=\varepsilon_i-\omega$ is the detuning of $i$th qubit of the cavity. Here, we choose to
drive the system by means of a small time-dependent quantities and separate them from the time-independent
system  denoting with superscript (0). As $g_i$ and $\Delta_i$ are both related to $\Phi_i$,
the time-dependent driven can be added by choosing the magnetic flux as $\Phi_i(t)=\Phi_i^{(0)}+\delta\Phi_i(t)$,
and the corresponding Hamiltonian can be written as $\hat{H}_{sub}=\hat{H}^{(0)}+\delta\hat{H}(t)$.
Assuming that the time-dependent fluxes $\delta\Phi_i(t)$ oscillate with the frequencies $\omega_i/2\pi$, the corresponding quantities are written as
\begin{subequations}
\label{parameter}
\begin{equation}
\delta\Phi_{i}(t)=F_{i}(t)\cos[{\omega}_{i}t+{\varphi}_{i} ],
\end{equation}
\begin{equation}
\delta\Delta_i(t)=L_i(t)\cos[\omega_it+{\varphi}_{i} ],
\end{equation}
\begin{equation}
{\delta}g_i(t)=T_i(t)\cos[\omega_it+{\varphi}_{i} ],
\end{equation}
\end{subequations}
where amplitudes $L_i(t)$ and $T_i(t)$ are determined by means of externally  modulated flux
amplitudes $F_i(t)$ based on equation (\ref{varepsilon}) and (\ref{g}).

The eigenvalues of the main Hamiltonian {$\hat{H}^{(0)}$} in the one excitation subspace
are $\{0,{\Delta}_1^{(0)},{\Delta}_2^{(0)},{\Delta}_3^{(0)}\}$.  In this eigenbasis, choosing
${\delta}g_i(t)$ and $\delta\Delta_i(t)$ to oscillate with frequency $\omega_i=\Delta_i^{(0)}$,
the effective Hamiltonian in rotating frame with rotating wave approximation reads \cite{tripod}
\begin{equation} \label{tri}
\hat{H}=\hbar\begin{bmatrix}
0&\Omega_1&\Omega_2&\Omega_3\\\Omega_1^*&0&0&0\\\Omega_2^*&0&0&0\\\Omega_3^*&0&0&0
\end{bmatrix},
\end{equation}
where effective Rabi frequencies are
 \begin{equation}
 \label{Rabi}
 \Omega_i=\eta_i L_i(t) e^{i\varphi_i(t)}
\end{equation}
with time-independent  parameter $\eta_i=g_i^{(0)}/(4\varepsilon_i^{(0)})-g_i^{(0)}/(2\Delta_i^{(0)})$.

To detect the  noncommutative feature of the non-Abelian GP, we now parameterize Rabi frequencies in Eq. (\ref{Rabi})
to form two specific evolution loops $C_1$ and $C_2$ with $U_1$ and $U_2$ being their respective evolution operators.
The  non-Abelian nature of GP is verified by the fact that  for an initial state
undergoing the two specific loops in different order will result in different final states. This noncommutative feature of the gauge structure
leads to the non-Abelian characteristic of the  non-Abelian GP.
For convenience,  the initial phase of Rabi frequencies are chosen as
 $\varphi_1 =\varphi_2 =0,\varphi_3=\xi$, respectively. Therefore, only the amplitudes of Rabi
frequencies vary with time, which are determined by the amplitudes $L_i(t)$ of the external magnetic flux.
Modulate $L_i(t)$ appropriately so that the two loops $C_1$ and $C_2$ are obtained as
\begin{subequations}
\begin{eqnarray}
C_1:&& \Omega_1=\Omega_0f(t),\notag\\ &&\Omega_2=\Omega_0f^2(t),\notag\\ &&\Omega_3=\Omega_{0}e^{-t^2/\tau^2}e^{i\xi},\\
C_2:&& \Omega_1'=\Omega_0f(t),\notag\\ &&\Omega_2'=\alpha\Omega_0f^2(t),\notag\\ &&\Omega_3'=\Omega_{0}e^{-{(t-\beta\tau)}^2/\tau^2}e^{i\xi},
\label{c2}
\end{eqnarray}
\end{subequations}
where $f(t)=\cos[{\pi t}/(2\tau)]$ for an interval of $t\in [-\tau, \tau]$ and  $\xi={\pi}t/\tau$.
Two variables $\alpha$ and $\beta$ make a distinction between the loops $C_1$ and $C_2$ with $\beta$  being a time delay factor.
At practical parametrization,  $\Omega_2= \Omega_0f^2(t)$ in the loop $C_1$.
To form $\Omega_2'$ in loop $C_2$, we introduce another magnetic flux which will produce $\Omega_4=(\alpha-1)\Omega_0f^2(t)$,
which is turned on only when forming the loop $C_2$.

Then,  Rabi frequencies $\Omega_i$ in $C_1$ can be rewritten as
\begin{eqnarray}
&& \Omega_1={\Omega}\sin{\theta}\cos{\phi}e^{i\xi_1}, \notag\\
&&\Omega_2={\Omega}\sin{\theta}\sin{\phi}e^{i\xi_2}, \notag\\
&& \Omega_3={\Omega}\cos{\theta}e^{i\xi_3}
\end{eqnarray}
with
\begin{eqnarray}
\Omega&=&\sqrt{|\Omega_1|^2+|\Omega_2|^2+|\Omega_3|^2}, \notag\\
\tan\theta(t)&=&\sqrt{({|\Omega_1|^2}+{|\Omega_2|^2})/|\Omega_3|^2}\notag\\
&=&\sqrt{({\cos^2({\pi}t/2\tau)}+{\cos^4({\pi}t/2\tau)})/{\exp}{(-2t^2/\tau^2)}}\notag\\
\tan\phi(t)&=&|\Omega_2|/|\Omega_1|=\cos({\pi}t/2\tau).
\end{eqnarray}
Then, two dark eigenstates of $\hat{H}$ in Eq. (\ref{tri}) are
\begin{eqnarray}
|D_1\rangle &=&\sin{\phi}e^{i\xi}|1\rangle-\cos{\phi}e^{i\xi}|2\rangle,\\
|D_2\rangle &=&\cos{\theta}\cos{\phi}e^{i\xi}|1\rangle+\cos{\theta}\sin{\phi}e^{i\xi}|2\rangle-\sin{\theta}|3\rangle, \notag
\end{eqnarray}
where $\{|1\rangle, |2\rangle, |3\rangle\}$ denote
$\{|0egg\rangle, |0geg\rangle, |0gge\rangle\}$, respectively. We then can get the gauge potential $A$  based on
$A_{ij\mu}={\langle}D_i|\frac{\partial}{\partial\chi^\mu}|D_j\rangle$  as  \cite{JGP}
 \begin{eqnarray}
&&A_{1,\theta=0}, \notag\\
&&A_{1,\phi}= \begin{bmatrix}
0&-\cos\theta\\\cos\theta&0
\end{bmatrix}, \notag\\
&&A_{1,\xi}=\begin{bmatrix}
i&0\\0&i\cos^2\theta
\end{bmatrix}.
\end{eqnarray}
Therefore, $A_1$  is
\begin{equation}
A_1=i(\frac{1+\cos^2\theta}{2}I+\frac{\sin^2\theta}{2}\sigma_z)d\xi-i\sigma_y{\cos}{\theta}d\phi,
\end{equation}
and its corresponding time evolution operator is
 \begin{equation}
U_1=\mathcal{P}\exp(-\int{A_{1,\mu}d\chi^\mu}),
\end{equation}
where $\mathcal{P}$ denotes the path-order operator.
In order to unambiguously detect non-Abelian  geometric phase, we confine
parameters $(\theta,\phi,\xi)$ vary from $(0,0,-\pi)$ to $(0,0,\pi)$
with time $-\tau\rightarrow\tau$. Similar to loop $C_1$, we can get the corresponding evolution operator $U_2$ based
on Eq. (\ref{c2}) in the loop $C_2$.

\begin{figure}[tbp]
\includegraphics[width=8cm]{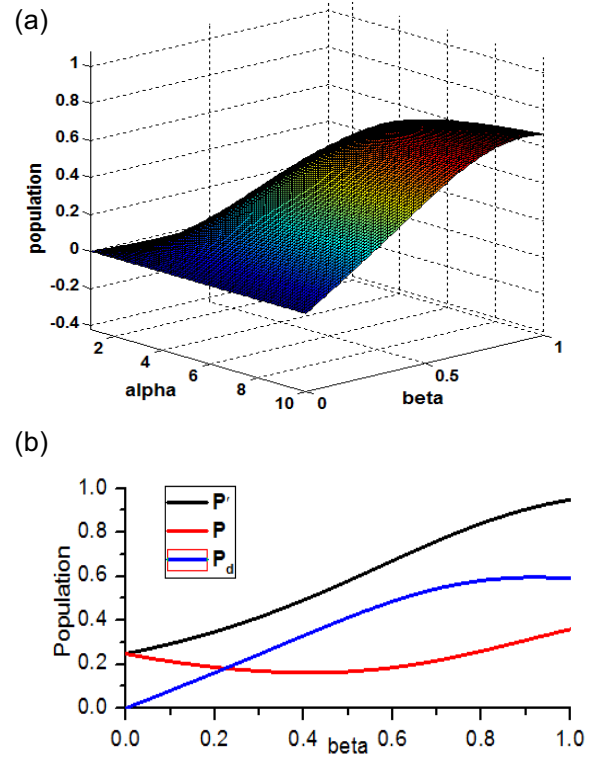}
\caption{(Color online) (a) Schematic of the population difference $P_d$ varying with parameters $\alpha$ and $\beta$.
(b) The populations $P$ (red), $P^\prime$ (black)
and $P_d$ (blue) as functions of the parameter $\beta$ $(0\rightarrow1.0)$ for the parameter $\alpha=6$. }
\label{jieguo}
\end{figure}

To detect the non-Abelian nature, we first prepare the
initial state as $|\psi\rangle_i=|D_2\rangle_i=|1\rangle$, and then
let it undergo two closed paths $C_1$ and $C_2$ in different orders, i.e.,
$U=U_2U_1$ (first $C_1$ then $C_2$) and $U^\prime=U_1U_2$ (first $C_2$
then $C_1$).  In order to implement the evolution $U (U^\prime)$, let $U_1$ ($U_2$)
in effect during time $-\tau\rightarrow\tau$, while $U_2$ ($U_1$)   during time $\tau\rightarrow3\tau$. The final states will  be
$|\psi\rangle_f=U|1\rangle=-U_{21}|2\rangle+U_{22}|1\rangle$,
$|\psi\rangle^\prime_f=U^\prime|1\rangle=-U^\prime_{21}|2\rangle+U^\prime_{22}|1\rangle$, respectively.
Therefore, the population difference  $P_d$  of the   two different
final states in $|1\rangle$ is
\begin{equation}
P_d=P^\prime-P=|U^\prime_{22}|^2-|U_{22}|^2.
\end{equation}
Whenever the $P_d \neq 0$ is detected, the noncommutative  feature of the non-Abelian GP is verified.
The population difference $P_d$ is numerically calculated with variables  $\alpha$ and $\beta$,
as shown in {Fig. \ref{jieguo} (a), which obviously indicates that $P_d \neq 0$}.
Fig. \ref{jieguo} (b) is a specific plot of the population difference $P_d$ with  $\beta$
as the only variable while $\alpha=6$, which shows maximum $P_d\approx 0.6$ when $\beta=0.9$.

Detecting the population difference of state $|1\rangle\equiv|0egg\rangle$ means that we just
need to observe population difference on the excited state $|e\rangle$ of qubit 1,
which can be realized by quantum non-demolition (QND) measurement. This can be achieved by tuning
the qubit dispersively coupled to the cavity with a large detuning $\Delta$,
and then measuring Hamiltonian will be  $\hat{H}_M=\hbar(\omega+\chi \sigma_z)a^{\dag}a$
with $\chi=g^2/\Delta$. We can then get a different frequency shift $\pm\chi$ of the
cavity mode with the qubit state on $|g\rangle$ and $|e\rangle$, respectively.
With  the coupling strength $g_1/2\pi=100$ MHz and the detuning $\Delta=5g$,
we can get the frequency shift as $\chi/2\pi=20$ MHz,
which is readily resolvable experimentally with high fidelity \cite{ARA}.


In summary, we have proposed an experimentally feasible scheme to detect the noncommutative
feature of the non-Abelian GP with effective tripod Hamiltonian in circuit QED.
The  non-Abelian nature of GP is verified by the fact that  for an initial state
undergoes the two specific loops in different order will lead to different final states.
This differences is detected through observing the population difference of state $|0egg\rangle$, which is achieved QND measurement in circuit QED.

\bigskip


This work was supported by the NFRPC (No. 2013CB921804),
the NSFC (No. 11004065), the PCSIRT,
the NSF of Guangdong Province,
and the Program of the Education Department of Anhui Province (No. KJ2012B075).

\end{document}